\documentclass{LT23auth}
\usepackage{graphicx}

\begin{document}

\begin{frontmatter}

\title{Spontaneous spin current near the interface between ferromagnets 
and unconventional superconductors}

\author[address1]{ Kazuhiro Kuboki\thanksref{thank1}},
\author[address1]{Hidenori Takahashi},

\address[address1]{Department of Physics, Kobe University, Kobe 657-8501, Japan }
\thanks[thank1]{ E-mail: kuboki@phys.sci.kobe-u.ac.jp}

\begin{abstract}
Proximity effects between ferromagnets (F) and superconductors (S) with 
broken time-reversal symmetry ($T$) are studied theoretically.  
For the S side we consider a chiral $(p_x\pm ip_y)$-wave,  and a 
$d_{x^2-y^2}$-wave superconductor,  the latter of which can form  
$T$-breaking surface state, i.e., $(d_{x^2-y^2} \pm is)$-state. 
The Bogoliubov de Gennes equation which describes the spatial variations 
of the superconducting order parameter and the magnetization is derived 
and solved numerically. 
It is found that a spontaneous spin current flows along the interface between the   
$(p_x \pm ip_y)$-wave superconductor and the ferromagnet. 
On the contrary, in the case of a  [110] interface of the $d_{x^2-y^2}$-wave SC,  
the surface state has a $(d \pm p_x \pm p_y)$-wave 
(or $(d_{x^2-y^2} \pm is)$-wave) symmetry,  
and thus no (only charge) spontaneous current arises. 
\end{abstract}

\end{frontmatter}


Recently  proximity effects between superconductors (S) and ferromagnets (F) attract
much attention,\cite{layer} especially when S side is an unconventional 
superconductor. In this paper we examine these effects theoretically. 

We consider two-dimensional S/F bilayer models at zero temperature.
The Hamiltonian for each layer is given by 
\begin{equation} 
\begin{array}{rl}
  H_{\rm L} & =  - \displaystyle t_{\rm L} \sum_{<i,j> \sigma} 
    ( c_{i,\sigma}^{\dagger} 
      c_{j,\sigma} + h.c.)  
     +  U_{\rm L} \sum_i n_{i\uparrow}n_{i\downarrow}  \\ 
     - & V_{\rm L}  \displaystyle 
     \sum_{<i,j>} \big[n_{i\uparrow}n_{j\downarrow}  
     + n_{j\uparrow}n_{i\downarrow}  \big], 
     \ \ ({\rm L} = {\rm S, F}) 
\end{array}     
\end{equation}
where $\langle i,j \rangle$ and $\sigma$ denote the nearest-neighbor 
pairs on a square lattice and the spin index, respectively. 
Parameters $t_{\rm L}$, $U_{\rm L}$ and $V_{\rm L}$ are the 
transfer integral, the on-site repulsive and the nearest-neighbor attractive 
interactions, respectively, for the ${\rm L} (={\rm S}, {\rm F})$ side.  
The transmission of electrons at the interface is described by the 
following tunneling Hamiltonian:
$
    H_{\rm T} =  -  t_{\rm T} \sum_{<l,m>\sigma} 
    ( c_{l,\sigma}^{\dagger} 
      c_{m,\sigma} + h.c.),      
$
where $l$ ($m$) denotes the surface sites of S (F) layer.
Then the total Hamiltonian of the system is $H = H_{\rm S} + H_{\rm F} + H_{\rm T} 
- \mu \sum_{i\sigma} c_{i\sigma}^\dagger c_{i\sigma}$ with 
$\mu$ being the chemical potential.  
The interaction terms are decoupled within the Hartree-Fock approximation:
$
U n_{i\uparrow}n_{i\downarrow} \to 
U\langle n_{i\uparrow} \rangle n_{i\downarrow} 
+ U\langle n_{i\downarrow} \rangle n_{i\uparrow}
-U \langle n_{i\uparrow} \rangle \langle n_{i\downarrow} \rangle, 
Vn_{i\uparrow}n_{j\downarrow} \to 
V\Delta_{ij}c_{j\downarrow}^\dagger c_{i\uparrow}^\dagger 
+V\Delta_{ij}^{*}c_{i\uparrow} c_{j\downarrow}
-V\vert \Delta_{ij} \vert^2 
$
with $\Delta_{ij} \equiv \langle c_{i\uparrow}c_{j\downarrow}\rangle$. 
The magnetization $m_i =\langle n_{i\uparrow} - n_{i\downarrow} \rangle/2$ 
and $\Delta_{ij}$ are the order parameters (OP's) 
to be determined self-consistently. 
The direction perpendicular (parallel) to the 
interface is denoted as $x$ ($y$), and we assume that the system 
is uniform along the $y$-direction.  
Along the $y$-direction we carry out the Fourier transformation, 
and the Bogoliubov de Gennes (BdG) equation which describes the 
$x$-dependences of OP's is derived and solved numerically.  
The system size we treat is $N_x = N_y =120$, and we take 
$t_{\rm S} = t_{\rm F} =1$ as the unit of energy. 
The parameters $U_{\rm L}$, $V_{\rm L}$ and $\mu$ are chosen such that 
the superconducting (SC) and the ferromagnetic order occurs in each 
layer. \cite{Hirsh,Mic,KK}
Depending on the electron density, $d_{x^2-y^2}$-wave, 
extended $s$-wave and  chiral $(p_x + ip_y)$-wave 
superconducting (SC) states are stabilized.

\begin{figure}[t]
\begin{center}\leavevmode
\includegraphics[width=0.9\linewidth]{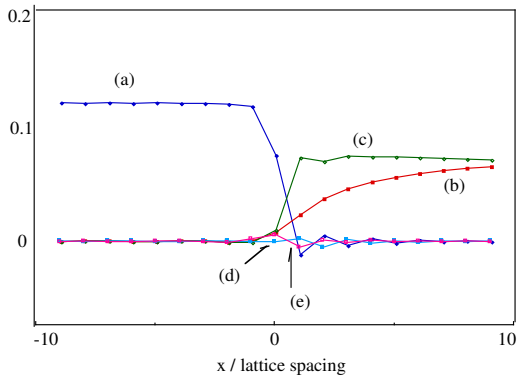}
\caption{Spatial variations of order parameters. The region $x < 0$  is a 
ferromagnet. 
(a) $m$, (b) Re$\Delta_{px}$, (c) Im$\Delta_{py}$, (d) Re$\Delta_d$ and 
(e) Re$\Delta_s$.  Note all OP's are non-dimensional. 
Parameters used are $t_{\rm T} =1, U_{\rm F} = 12,
U_{\rm S} = 1.5, V_{\rm S} =2.5, V_{\rm F} =0$ and $\mu =-1.6$.
}
\label{figurename}\end{center}\end{figure}

Here we  consider $d_{x^2-y^2}$-wave  and  $(p_x + ip_y)$-wave SC states 
for the S layer. 
It is known that near the [110] surface of the former the system may break 
time-reversal symmetry ($T$) by introducing second component of SCOP 
$(is)$ and that a 
spontaneous (charge) current flows along the surface. \cite{Sig}
The chiral SC states also break $T$ and the spontaneous current can flow 
along the edge of the system.\cite{matsu}
When  S and F layers are connected, the magnetization (spin-triplet SCOP) 
may be induced in the former (latter) due  to the proximity effect.\cite{KK2}
Then it may be possible to have a spontaneous spin current 
(defined as  $J_{\rm spin} (x) \equiv J_{y\uparrow}(x) - J_{y\downarrow}(x)$) 
along the interface of a $T$-breaking S layer and a F layer, 
because of the imbalance of the electron densities of spin-up and spin-down 
electrons. 

First we investigate the case of a bilayer composed of a 
chiral $(p_x \pm ip_y)$-wave 
superconductor and a ferromagnet. 
In Fig.1 we show the spatial variations of OP's for the case of [100] interface. 
(We have also studied the case of [110] surface, and the results were 
qualitatively the same.) 
It is seen that $p_x$ and $p_y$-wave components (denoted as 
$\Delta_{px}$ and $\Delta_{py}$, respectively) are suppressed near 
the interface,  while $d$-wave  ($\Delta_d$) and $s$-wave ($\Delta_s$) 
components are induced.  
The magnetization penetrates into the S layer and it coexists with SCOP's. 
We can see that a spontaneous spin current actually flows (Fig.2). 

Next we  consider the $d_{x^2-y^2}$-wave SC state.
We consider the [110] interface, since a $T$-breaking 
surface state associated with a spontaneous current can be formed in this case, 
but not in the case for a [100] surface. 
The results are qualitatively different depending on the 
value of $t_{\rm T}$. 
When $t_{\rm T}$ is comparable to $t_{\rm S}$ 
($= t_{\rm F} =1$) , the magnetization penetrates into the S layer 
and then the $p$-wave components are induced. 
However, the complex component ($is$) is not induced in contrast to 
the case of a [110] surface exposed to the vacuum. 
The resulting state has a $(d_{x^2-y^2} \pm p_x \pm p_y)$-symmetry, 
and no spontaneous current arises. 
When the value of $t_{\rm T}$ is reduced the magnetization does not 
penetrate into the S layer (and thus $p$-wave OP's are absent), 
and the state has a $(d_{x^2-y^2} \pm is$)-symmetry. 
In this case a spontaneous charge current flows along the interface,  
but not the spin current because the electron densities 
of spin-up and spin-down electrons are  the same. 
The transition between the above two states seems to be of first order, 
and we did not find states with spontaneous spin currents. 

In summary we have studied the states near the interface between unconventional 
superconductors and ferromagnets. Spontaneous spin currents are 
found in the case of chiral $(p_x\pm ip_y)$-wave superconductors. 
For $d_{x^2-y^2}$-wave superconductors the resulting states have   
either $(d_{x^2-y^2}\pm is)$- or 
$(d_{x^2-y^2} \pm p_x \pm p_y)$ -symmetry, so that 
only charge (no) current appears in the former (latter).

\begin{figure}[t]
\begin{center}\leavevmode
\includegraphics[width=0.9\linewidth]{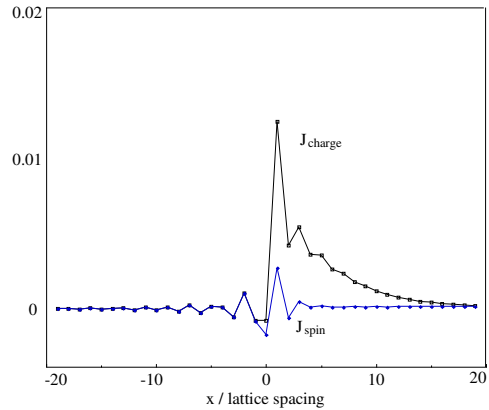}
\caption{Spontaneous charge and spin currents along the interface,
in units of $eta/\hbar$ ($a$ being the lattice constant).  
}
\label{figurename}\end{center}\end{figure}

  
%
%
\medskip
The authors  thank M. Sigrist, H. Shiba and H. Fukuyama for useful discussions. 

%
%


\begin{thebibliography}{9}

\bibitem{layer} R Meservey and P.M Tedrow, Phys. Rep. {\bf 238} (1994) 173. 

\bibitem{Hirsh} J. E. Hirsh, Phys. Rev. B {\bf 31} (1985) 4403.

\bibitem{Mic} R. Micnas, J. Ranninger and S. Robaszkiewicz: 
Rev. Mod. Phys. {\bf 62} (1990) 113.

\bibitem{KK}  K. Kuboki:, J. Phys. Soc. Jpn. {\bf 70} (2001) 2698. 

\bibitem{Sig} For a review on $T$-breaking SC states, see for example, 
M. Sigrist, Prog. Theor. Phys. {\bf 99} (1998) 899. 

\bibitem{matsu} M. Matsumoto and M. Sigrist, 
J. Phys. Soc. Jpn. {\bf 68} (1999) 994. 

\bibitem{KK2} K. Kuboki, J. Phys. Soc. Jpn. {\bf 68} (1999) 3150. 

\end{thebibliography}
\end{document}